\documentclass[useAMS,usenatbib]{mn2e}
\usepackage{amssymb}
\usepackage{amsmath}
\usepackage{txfonts}
\usepackage{graphicx}
\usepackage{natbib}
\usepackage{rotating}
\usepackage{url}
\usepackage{epsfig}
 
\def\rxte{\textit {RXTE}}

\def\xmm{\textit {XMM-Newton}}
\def\xmmsp{\textit {XMM-Newton }}

\def\deg {$^\circ$}
\def\lsi {LSI~+61\deg~303}
\def\gr{$\gamma$-ray}

\usepackage{color}
\definecolor{red}{rgb}{0.7,0,0}
\definecolor{blue}{rgb}{0,0,0.7}

\begin{document}

\title{Compactified pulsar wind nebula model of  \gr
-loud binary \lsi.} 

 \author[A. Neronov and M. Chernyakova] {A. Neronov$^{1,2}$ \thanks{E-mail:Andrii.Neronov@obs.unige.ch} and 
M. Chernyakova$^{1,2}$\thanks{M.Chernyakova 
is on leave from Astro Space Center of the P.N.~Lebedev Physical Institute,  Moscow, Russia} \\
$^{1}$INTEGRAL Science Data Center, Chemin d'\'Ecogia 16, 1290 Versoix, Switzerland\\
$^{2}$Geneva Observatory, University of Geneva, 51 ch. des Maillettes, CH-1290 Sauverny, Switzerland}

\date{Received $<$date$>$  ; in original form  $<$date$>$ }

\pagerange{\pageref{firstpage}--\pageref{lastpage}} \pubyear{2005}

\maketitle

\label{firstpage}

\begin{abstract} 
We show that radio-to-TeV  properties of the binary system \lsi\ can  be
explained by interaction of the compact object (a young pulsar) with the
inhomogeneities of the wind from companion Be star. We develop a model 
scenario of "compactified"
pulsar wind nebula  formed in result of such interaction. To test the model
assumptions about geometry of the system we re-analyze
the available X-ray observations to study in more details the 
variations of the hydrogen column density on long (orbital) and short (several
kilosecond) time scales.
\end{abstract}


\section{Introduction}

The Be star binary \lsi\ is one of the three currently known \gr -loud
X-ray binaries. The spectrum of high-energy emission from the system
extends up to TeV energies \citep{albert06} and the power output of
the source is dominated by emission in the \gr\ energy band.

The origin of the high-energy activity of the source is not completely
clear.  The problem is that most of the "prototypical"
accretion-powered X-ray binaries do not reveal significant \gr\
activity. It is possible that the "\gr -loudness" of an X-ray binary
can be related to its special orientation with respect to the line of
sight (by analogy with the case of \gr -loudness of active galactic
nuclei) \citep{mirabel99}. Otherwise, it is possible that the three
X-ray binaries detected in \gr\ band so far are fundamentally
different from the conventional X-ray binaries. For example, the \gr
-loud binaries can be powered by a different mechanism, than the
conventional X-ray binaries. In fact, one of the three \gr -loud
binaries, PSR B1259-63, is known to be powered by the rotation energy
of a young pulsar, rather than by accretion onto a black hole or a
neutron star \citep{johnston92}. In two other \gr -loud X-ray
binaries, namely LS 5039 and \lsi, the pulsed emission from the
pulsar was not detected, so there is no direct evidence for the pulsar
powering the activity of these sources\footnote{The binary orbits of
  these two sources are much more compact than that of PSR
  B1259-63. In this case the radio pulsed emission is absorbed in the
  wind from companion star}. However, similarity of the spectral
energy distributions of the three systems enables to make a conjecture
about similar mechanism of their activity.

If the activity of \lsi\ is powered by a young pulsar, the
radio-to-\gr\ emission is generated in the course of collision of
relativistic pulsar wind with the wind from companion
star. Interaction of the pulsar and stellar winds leads to formation
of a "scaled down" analog of the pulsar wind nebulae (PWN) in which
the energy of the pulsar wind is released at the astronomical unit,
rather than on parsec, distance scale~\citep{neronov06}.

In this paper we explore the structure of compactified PWN of \lsi. We
show that different physical processes determine cooling of
high-energy particles at different distances from Be star. This leads
to an "onion-like" structure in which the region of dominance of
Coulomb losses is embedded into the region of dominance of inverse
Compton (IC) losses which is, in turn, situated inside the region of
dominant synchrotron loss. The density and inhomogeneity of the
stellar wind determine the speed of escape of the high-energy
particles injected in the region of pulsar/stellar wind
interaction. Anisotropy of the stellar wind leads to the dependence of
the escape speed on the orbital phase and, as a consequence, to the
variations of the relative importance of Coulomb, IC, and synchrotron
losses along the orbit. We show that such model explains the puzzling
behaviour of radio, X-ray and \gr\ lightcurves of the system (shifts
of the maxima from the periastron, shifts between the maxima of X-ray
and radio lightcurves, shifts of the maxima from orbit to orbit).

In order to better constrain our model we re-analyze the existing
X-ray observations of \lsi\ and find a marginally detected variations
of the hydrogen column density on long (orbital) and short
(kilosecond) time scales. The short time scale variations reveal the
small-scale inhomogeneities of the stellar wind which lead to the macroscopic
mixing of pulsar and stellar winds. If confirmed with more X-ray
observations at the orbital phases close to the periastron, the
detected orbital modulation of the hydrogen column density 
constrains the geometry of the system.

\section{Interaction of the pulsar wind with a clumpy wind from 
companion star.}   
\label{model}

\subsection{Basic properties of \lsi.}

Radio emission from the system is known to be periodic with the period
of $T=26.4960$~d which is associated with the binary orbital period
\citep{gregory02}.  Optical data allow to constrain the orbital
parameters of the system revealing the eccentricity of the orbit,
$e\simeq 0.55$ \citep{grundstrom06}. The binary orbit of \lsi\ turns
out to be very compact, with the periastron at just $1.3$ radii of Be
star (see Fig.  \ref{fig:scheme}).

The existing measurements are not sufficient to determine the nature
of the compact object (neutron star or black hole), because the
inclination of the orbit is poorly constrained.  This uncertainty
allows to discuss two possible models for the origin of the \lsi\
activity. Models of the first type, first introduced by
\citet{taylor84} (see \citet{boshramon06} for a recent reference), assume that activity of the source is powered by
accretion onto the compact object. In the second class of models,
first proposed by \citet{maraschi81}, the activity of the source is
explained by interactions of a young rotation powered pulsar with the
wind from the companion Be star. The fact that no pulsations from the
system has been found can be explained by the absorption of the pulsed
radio emission in the Be star wind. The absence of the break up to the
100 keV in the X-ray spectrum and similarity of the radio-to-TeV \gr\
spectral energy distribution of the source to the one of PSR B1259-63
favor the "hidden pulsar" model \citep{chernyakova06}.  The "hidden
pulsar" model is also favored by the recent VLBA monitoring of the
\lsi\, which reveals an extended source whose irregular morphology varies on
the orbital time scale \citep{dhawan06}.

\subsection{Basic model of pulsar/stellar wind interaction.}

In the most simple version, the model of interaction of the pulsar
wind with the wind from companion Be star assumes that an isotropic
relativistic outflow from the pulsar hits a homogeneous (but
anisotropic, in the case of Be star companion) outflow from the star
along a regular bow-shaped surface. Such a model was developed in
details e.g. in relation with the PSR B1259-63 by \cite{tavani97} and
applied to the case of \lsi\ by \cite{dubus06}.  Geometry of the
interaction surface is determined by the pressure balance between the
pulsar and stellar winds. In the settings of "homogeneous stellar
wind" scenario, the pulsar and stellar wind do not mix
macroscopically, which allows the shocked pulsar wind to escape from
the system with the speed $\sim 10^{10}$~cm/s, much higher than escape
velocity of the shocked stellar wind.

\begin{figure} 
\includegraphics[width=\linewidth]{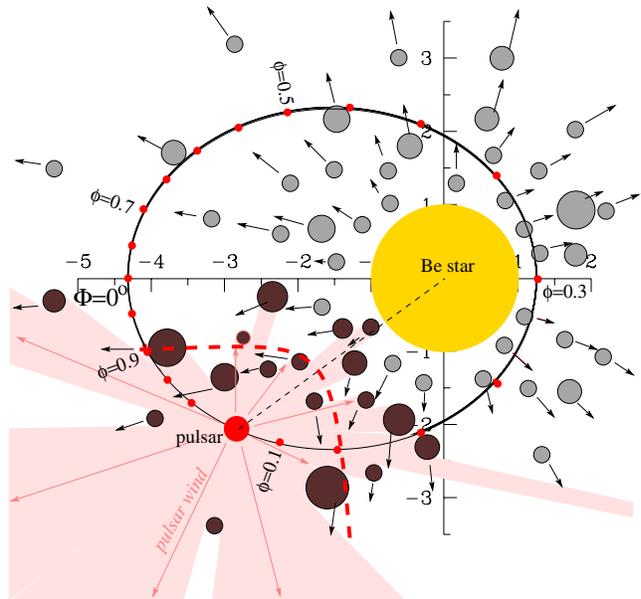}
\caption{Schematic representation of interaction of the pulsar wind with a
clumpy wind from companion Be star. Relativistic pulsar wind hits the nearest
clumps of the stellar wind. Instead of a smooth bow-shaped contact surface
(red dashed line) the interaction region (the set of the dark shaded
clumps) is irregularly shaped. Ellipse represents correctly scaled 
pulsar binary orbit around the companion star. $x$ and $y$ axes are in units of
the radius of companion star. Red dots along the orbit mark intervals regularly
spaced in phase, $\Delta\phi=0.05$. The periastron corresponds to $\phi=0.3$ 
\citep{grundstrom06}. }
\label{fig:scheme} 
\end{figure}

The assumption of homogeneity of the stellar wind is initially adopted
for the sake of simplicity of the model, rather than by physical
motivations. At the same time, it is known that intrinsic
instabilities of the winds from massive stars lead to formation of
large inhomogeneity of the winds, observationally seen in X-ray
emission from massive stars and in line-profile variability (see
e.g.  \citet{puls06}). Significant inhomogeneity of the stellar wind
leads, in general, to disappearance of the regular bow-shaped contact
surface of the pulsar and stellar winds and to a macroscopic mixing of
the two winds inside an irregularly shaped interaction region, as it
is shown in Fig. \ref{fig:scheme}.  The macroscopic mixing of the
pulsar and stellar winds slows down the escape of the shocked pulsar
wind.

Different regimes of escape of the high-energy particles injected into
the region of interaction of pulsar and stellar winds lead to
significantly different predictions about the spatial structure of the
compactified PWN in the "homogeneous stellar wind" and "clumpy stellar
wind" scenaria. The reason for this is the strong radial gradient of
the densities of matter and soft photons around the system. Depending
on the velocity of escape, high-energy electrons can loose different
fraction of their energy onto Coulomb, IC and synchrotron loss. Below
we explore in details the relative importance of different cooling
mechanisms and propose a model of the compact PWN of \lsi.

\subsection{Inhomogeneity of the stellar
 wind and short time scale variability of the source.}

Long \xmm\ observation of \lsi\ in 2005 has revealed variability of
the system at $\Delta t\simeq (several)$~ks time scale, which is much
shorter than the orbital period \citep{sidoli06}. Variability at a
similar time scale is observed also in radio energy band
\citep{peracaula97}.  The observed variability indicates that the
characteristics of interaction of the pulsar and stellar winds change
at  distance scales which are smaller than the size of the binary
orbit. Distance scale relevant for the fast variability can be
estimated as
 \begin{equation}
R_{w}\sim v\Delta t \sim 10^{11}\left[\frac{v}{10^7\mbox{ cm/s}}
\right]\left[\frac{\Delta t}{10\mbox{ ks}}
\right]\mbox{ cm}
\end{equation}
where $v$ is typical velocity scale (e.g. the orbital velocity of the 
pulsar, or the speed of the stellar wind). 

If one assumes that the power output in the pulsar wind is constant, 
the distance scale related to the fast variability has to be
associated with the pulsar wind interactions with inhomogeneities (clumps) of 
the stellar wind.  Changes in the X-ray luminosity can be caused by the 
variaitons of the number of
inhomogeneities exposed to the pulsar wind 
and/or variations of characteristics (density, size,
magnetic field strength) of inhomogeneities. 

As it is explained above, the inhomogeneity of the stellar wind leads 
the disappearence of a regular bow-shaped surface of pulsar/stellar wind
interaction. The irregularity of the contact surface of wind interaction
 leads to a change in the regime
of escape of high-energy particles from the vicinity of Be
star. Namely, contrary to the the "homogeneous wind" scenario, in
which the shocked pulsar wind escapes along the bow-shaped contact
surface in a time $t_{esc}\sim 3D/c\sim 30\left[10^{12}\mbox{ cm}
  /D\right]$~s (which depends on the distance $D$), the escape of the
pulsar wind mixed with the stellar wind can slow down to
\begin{equation} 
t_{esc}\sim D/v_{w}\sim
10^5\left[\frac{10^{12}\mbox{ cm}}{D}\right]\left[\frac{10^7\mbox{
cm/s}}{v_{wind}}\right]\mbox{ s} 
\end{equation} 
The wind from the Be star is known to have anisotropic structure with
a dense and "slow" wind ($v_{w}\sim 10^6\div 10^7$~cm/s) in the
equatorial region and rarefied "fast" wind ($v_{wind}\sim
(several)\times 10^8$~cm/s) from the polar regions. The difference in
the escape time changes the balance between the synchrotron, IC and
Coulomb energy losses of high-energy electrons injected in the region
of interaction of pulsar and stellar winds.

\subsection{Cooling of high-energy electrons in the clumps.}

Following \citet{maraschi81} and \citet{chernyakova06} we assume that
X-ray emission from the system at  energies $\epsilon_{IC}\sim
1-10$~keV is dominated by the IC scattering of optical
photons from the Be star by electrons of energies about
\begin{equation}
\label{eic}
E_{e,IC}\simeq 10\left[\frac{2\times 10^4\mbox{ K}}{T_*}\right]^{1/2}
\left[\frac{\epsilon_{IC}}{4\mbox{ keV}}\right]^{1/2}\mbox{ MeV}
\end{equation}
($T$ is the temperature of the Be star).  Radio flux at the frequency
$\nu_{synch}$ is produced via
synchrotron emission from electrons with roughly the same energies,
\begin{equation} 
\label{esynch}
E_{e,synch}\simeq
10\left[\frac{1\mbox{ G}}{B}\right]^{1/2}\left[ \frac{\nu_{synch}}{7\mbox{
GHz}}\right]^{1/2}\mbox{ MeV} 
\end{equation} 
($B$ is the magnetic field strength).  

10~MeV electrons are injected in the inhomogeneities of the stellar wind
 irradiated by the relativistic pulsar wind.  Such
injection can be either the result of shock acceleration of electrons
from the stellar wind or the result of cooling of higher energy
electrons from the pulsar wind.

The high-energy particles can be retained in inhomogeneities with strong
enough magnetic field.  Assuming that electrons diffuse in disordered
magnetic field, one can estimate the escape time (in the Bohm
diffusion regime) as
\begin{equation}
t_{diff}\simeq 3\times 10^6\left[\frac{R_{w}}{10^{11}\mbox{ cm}}\right]^2
\left[\frac{B}{1\mbox{ G}}\right]\left[\frac{10\mbox{ MeV}}{E_e}\right]\mbox{ s}
\end{equation}
Comparing the escape time with the inverse Compton cooling time
\begin{equation}
\label{eq:tic}
t_{IC}(D,E_e)\simeq 10^4\left[\frac{L_*}{10^{38}\mbox{ erg/s}}\right]
\left[\frac{D}{10^{12}\mbox{ cm}}\right]^2
\left[\frac{10\mbox{ MeV}}{E_e}\right]\mbox{ s}
\end{equation}
($L_*$ is the luminosity of Be star) and/or synchrotron cooling time, 
\begin{equation}
\label{eq:tsynch}
t_{synch}\simeq 3\times 10^5\left[\frac{1\mbox{ G}}{B}\right]^2
\left[\frac{10\mbox{ MeV}}{E_e}\right]\mbox{ s}
\end{equation}
one  finds that if the magnetic field in the clump is $B_{clump}\sim
1$~G, then electrons captured in the clumps can efficiently cool before
they escape.

The binary orbit of \lsi\ is very compact, so that at periastron the
pulsar approaches the Be star as close as 1.3 stellar radii, $D\simeq
5\times 10^{11}$~cm \citep{grundstrom06}.  Close to the surface of the
star, not only the density of photon background is high, but also the
density of the stellar wind, $n$. In this case 10~MeV electrons suffer
from the strong Coulomb energy loss. The Coulomb loss time,
\begin{equation}
\label{eq:tcoul}
t_{Coul}\simeq 2\times 10^3\left[\frac{10^{10}\mbox{ cm}^{-3}}{n}\right]
\left[\frac{E_e}{10\mbox{ MeV}}\right]\mbox{ s}
\end{equation}
becomes shorter than the IC cooling time (\ref{eq:tic})
below the "Coulomb break" energy
\begin{equation}
E_{e,Coul}\simeq 20\left[\frac{L_*}{10^{38}\mbox{ erg/s}}\right]^{1/2}
\left[\frac{D}{10^{12}\mbox{ cm}}\right]
\left[\frac{n}{10^{10}\mbox{ cm}^{-3}}\right]^{1/2}\mbox{ MeV}
\end{equation}
Most of the power injected in the form of electrons with
energies below $E_{e,Coul}$ goes into heating of the stellar wind, rather
than into the synchrotron and IC emission. If the density of the wind
is
\begin{equation}
\label{n}
n>10^{10}\left[\frac{10^{12}\mbox{ cm}}{D}\right]^2\mbox{ cm}^{-3}
\end{equation}
both the X-ray emission in the energy band $\epsilon_{IC}<10$~keV and
radio emission below $\nu_{synch}\sim 10$~GHz (corresponding to
$E_e\simeq 20$~MeV) are suppressed. In other words, the
innermost part of the stellar wind is "X-ray/radio dim".

Radial gradients of the densities of the stellar wind and of the soft
photon background lead to significant variations of the relative
importance of Coulomb, IC and synchrotron energy losses with
distance. This, in turn, leads to a complicated radial structure of
the compact PWN.

\subsection{Structure of the compactified pulsar wind nebula.}

Recent VLBA monitoring of the source over an entire orbital cycle
\citep{dhawan06} reveals an extended radio source of a variable
morphology with the overall size of $D_{PWN}\sim (several)\times
10^{14}$~cm which can be identified with a "compactified" PWN, similar
to the compactified PWN of PSR B1259-63 \citep{neronov06}. The radio
emission at 3-13~cm wavelengths (synchrotron photon energies
$\epsilon_{synch}\sim (1\div 3)\times 10^{-5}$~eV) peaks outside the
binary orbit, at the distance $D_r\sim (several)\times
10^{13}$~cm. The position of the peak moves 
around the central Be star on the orbital time scales.

The variability of morphology of the source on the orbital time scale
shows that the cooling time of 10~MeV electrons responsible for the
radio synchrotron emission is $t_{PWN}\le 10^6$~s. Comparing this time
with synchrotron and inverse Compton cooling times in the region of
the size $D_{PWN}$ one can find that $t_{PWN}$ is shorter than the
inverse Compton cooling time, $t_{IC}(D_{PWN}, 10\mbox{ MeV})$
(\ref{eq:tic}). Equating $t_{PWN}\simeq t_{synch}$ (\ref{eq:tsynch})
one can find the magnetic field strength in the compactified PWN,
$B_{PWN}\simeq 1\mbox{~G}$.

Suppression of the radio synchrotron emission from the region $D<D_r$ 
can be explained by the fact that in the inner part of the
PWN the inverse Compton energy loss dominates over the synchrotron loss.
Indeed, from (\ref{eq:tic}), (\ref{eq:tsynch}) one can find that 
if the magnetic field in the nebula does not rise significantly toward the
center, $t_{IC}< t_{synch}$ for the distances $D<10^{13}$~cm.
Thus, the inverse Compton emission comes predominantly from the inner part of
the nebula, while the synchrotron emission mostly comes from its outskirts.

High-energy electrons responsible for the radio and X-ray emission are
initially injected in the region of interaction of the pulsar and
stellar wind.  Taking into account that the binary orbit of \lsi\ is
very compact, $D_{orb}\sim (several)\times 10^{12}$~cm, one finds that
the high-energy electrons can fill the entire compactified nebula
before they loose all their energy on synchrotron emission only if
they spread from the compact injection region with the speed $v\ge
D_{PWN}/t_{synch}\simeq 10^8\mbox{ cm/s}$. There are two possibilities
for such "fast" escape. Either electrons travel with the nearly
relativistic speed in the shocked pulsar wind, so that $v\sim
10^{10}$~cm/s, or they are injected into the nebula during the periods
when the pulsar interacts with the clumps of the fast polar wind from
Be star, which has the velocity $v_{wind}\ge 10^8$~cm/s.

In the former case it would be difficult to explain why the IC
luminosity of the system is much higher than the synchrotron
luminosity. Indeed, electrons would leave the compact region of the
size $D<10^{13}$~cm in less than one kilosecond. This time is shorter than
the IC cooling time and in such scenario most of the power of high
energy electrons would be released via synchrotron, rather than via IC
emission.

To the contrary, if electrons escape with the speed of the stellar
wind, the periods of fast escape in the polar wind are intermittent
with the periods of slow escape in equatorial wind. The velocity of
the equatorial wind is $v_{wind}\sim 10^6\div 10^7$~cm/s. It takes
electrons some $10^6$~s to escape to the distances beyond $10^{13}$~cm
during the pulsar propagation through the equatorial wind. Since the 
escape time is larger than the IC cooling time in this case,
most of the power is released via IC X-ray emission.

It is clear from the above discussion that the X-ray IC emission from
the system is produced in the compact region, $D\le 10^{13}$~cm, in
the direct vicinity of the Be star. However, close to the stellar
surface, electrons can suffer from the severe Coulomb loss. Since the
radial profile of the density of the stellar wind in the equatorial
disk, $n\sim D^{(-2)\div (-4)}$ (e.g. \citet{waters88}), is steeper than the soft photon
density profile, $n_{ph}\sim D^{-2}$, the Coulomb loss can dominate
over the IC loss in the innermost part of the compactified PWN. The
X-ray IC emission from this innermost part is suppressed and most of
the energy of 10~MeV electrons goes into heating of the stellar wind
via Coulomb collisions.

\begin{figure}
\includegraphics[width=\linewidth]{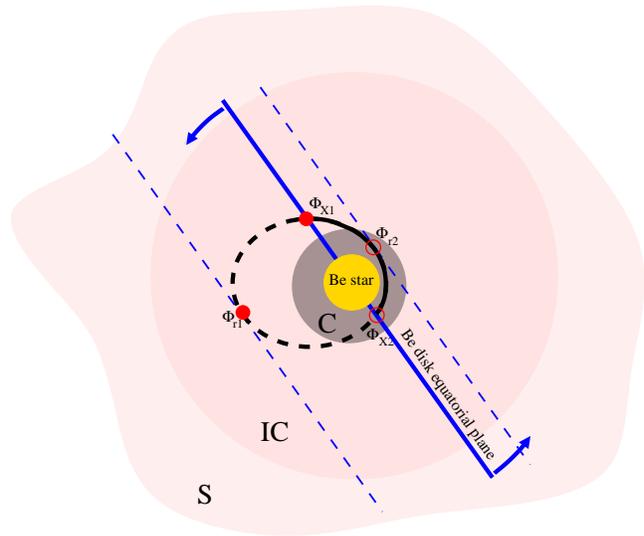}
\caption{Onion-like structure of the compactified pulsar wind nebula
  of \lsi.  The Coulomb losses dominate in the innermost region marked
  as "C". The inverse Compton loss dominates in the region marked
  "IC". The synchrotron loss dominates in the outer region marked as
  "S". The ellipse shows the binary orbit. The suggested orientation
  of the equatorial plane of the geometrically thick Be star disk at
  the moment of 1996 \rxte\ observations is shown by the thick solid
  blue line. The part of the binary orbit behind the equatorial plane
  of Be star disk is
  shown by the dashed line. $\Phi_{X1}$ and $\Phi_{r1}$ mark the phases of maxima of X-ray and radio lightcurves (see text). }
\label{fig:scheme2}
\end{figure}

Thus, the entire compactified PWN has the "onion-like" structure shown
in Fig.  \ref{fig:scheme2}: in the innermost region of the nebula the
dominant Coulomb loss suppresses both the IC and synchrotron
luminosity, in the intermediate region $10^{12}$~cm$<D<10^{13}$~cm the
IC energy loss dominates while in the outer part of the nebula
$10^{13}$~cm$<D<(several)\times 10^{14}$~cm the synchrotron loss
dominates.

\subsection{X-ray and radio lightcurves of the source.}

X-ray (and radio) lightcurve of the system exhibit a single maximum
per orbit, see Fig. \ref{lc}. In this figure we plot the X-ray flux
as a function of the "angular" orbital phase, $\Phi$, defined in such
a way that the apastron is at $\Phi=0^\circ$, while the periastron is
at $\Phi=180^\circ$. The phase of the maximal X-ray flux is known to
be shifted with respect to the maximum in radio band. Besides, the
phases of X-ray and radio maxima shift from orbit to orbit.  The
overall variation of the X-ray flux from the system is by a factor of
2-3.  The radio flux shows larger variations, by a factor of 10. The
onion-like structure of the compactified PWN discussed in the previous
subsection gives a clue to understand the behaviour of the X-ray and
radio lightcurves.

The X-ray flux increases when the pulsar moves deeper into the slow
equatorial wind of Be star.  During these periods the high-energy
electrons responsible for the X-ray emission are retained for a longer
time in the vicinity of Be star where they efficiently loose energy
via IC scattering. In such a model the maximum of the X-ray lightcurve
at the phase $\Phi_{X1}\simeq 300^\circ$, observed with RXTE, has to be
identified with the moment of the pulsar passage through the
equatorial plane of the Be star disk (in 1996). This fixes the
orientation of the disk is fixed as it is shown in
Fig. \ref{fig:scheme2}.  

Orbit-to-orbit fluctuations/shifts of the position of
X-ray maximum are expected because the pulsar passes through the
innermost part of the disk and the pulsar wind is injected into
several individual clumps in the disk, rather than into the "average"
equatorial wind of Be star. Drift of the phase of X-ray maximum,
$\Phi_{X1}$ on 4.6~yr year time scale of variability  of Be star disk
\citep{zamanov00} is expected if e.g. the disk precesses or a
significant asimutal inhomogeneity of the disk changes its
orientation.

By analogy with the case of PSR B1259-63, one expects to find a second
maximum of the X-ray lightcurve at the phase $(\Phi_{X2}=\Phi_{X1}-180^\circ\simeq
120^\circ$ (assuming the 1996 disk position), during the
pre-periastron passage of the pulsar through the equatorial plane of
Be star disk. However, during the second disk passage the pulsar is
much closer to the surface of Be star, where the Coulomb loss can
dominate over the IC loss. In this case most of the illuminated clumps
of Be star wind are "X-ray dim", as it is explained in the previous
sub-section. The decrease of amount of the X-ray bright clumps leads
to the suppression of the second maximum of the X-ray lightcurve.

It is clear from the previous subsection that 10~MeV electrons can
fill the entire compact PWN and efficiently emit synchrotron radiation
in radio band only if they escape from the compact injection region
with the speed $v\ge 10^8$~cm/s, much larger than the velocity of the
equatorial wind. This leads to a suggestion that the radio luminosity
of the system peaks during the period when the pulsar wind can be
efficiently injected into the clumps of the polar wind of Be
star. These period(s) correspond to the orbital phase when the pulsar
moves to its highest position above or below the equatorial plane. (If
the binary orbit would be circular, the phase of the radio maximum
would be $\Phi_{r1}\simeq\Phi_{X1}+90^\circ$. The eccentricity of the orbit
shifts the highest elevation above the disk to earlier or later phase. 
The phase of the radio maximum is found by drawing a tangent  
to the ellipse parallel to the assumed line of intersection of the orbital plane with the equatorial plane 
of Be star disk (the dashed lines in Fig. \ref{fig:scheme2}). Similarly to the case of
X-ray lightcurve, the second maximum of the radio lightcurve, naively
expected at $\Phi_{r2}$ (see Fig. \ref{fig:scheme2}), is missing because of the
close approach of the binary orbit to the surface of Be star.

The phase of the maximum of the radio lightcurve is known to exhibit a
periodic drift on the 4.6~yr timescale \citep{gregory02}. Within the
proposed model, the drift of the radio maximum is associated to the
drift of the phase of the highest elevation of the binary orbit above
the equatorial plane of the Be star disk or, equivalently, of the
orientation of the line of intersection of the equatorial plane of the
Be star disk with the orbital plane. Such drift of the orientation of the 
line of intersection can be e.g. due to the precession of the disk with the period of 4.6~yr. 

It is interesting to note that the pattern of the systematic 4.6~yr drift of the
phase of the radio maximum agrees well with the assumption of the
change of the mutual orientation of the orbital plane and the
equatorial plane of Be star disk. Suppose that the line of
intersection of the binary orbit with the equatorial plane of Be star
disk rotates with a period of $P\simeq 2\times 4.6$~yr$\simeq
9.2$~yr. Suppose that at an initial moment the line of intersection 
of equatorial plane of Be star disk coincides with the major axis of the binary orbit. At this 
moment the phase of the maximal elevation of the binary orbit above the disk is roughly equal to $\Phi_{r1}\sim 290^\circ$
or $\Phi_{r1}\simeq 70^\circ$
(for the particular case of the binary orbit of \lsi, see Fig. \ref{fig:scheme} and \ref{fig:scheme2}.  
If the line of intersection of the orbital plane with the equatorial plane of Be star disk precesses in the counter-clockwise 
direction, the phase of the highest elevation of the orbit
above the equatorial plane of the disk grows from $\sim 290^\circ$ to $360^\circ$ and then further from $
0^\circ$ to $\simeq 70^\circ$. In terms of the orbital phase,
the phase of radio maximum changes from 0.45 to 1. and then from 0. to
0.1. At the moment when the line of intersection of orbital and
equatorial plane again coincides with the major axis of the orbit, the phase
of the radio maximum should make a ``flip'' from $\Phi_{r2}\simeq
70^\circ$ to $\Phi_{r1}\simeq 290^\circ$ (from $\phi\simeq 0.1$ to
$\phi\simeq 0.45$). Since the line of intersection of orbital and
equatorial planes coincides with the major axis twice per period, the
``flips'' of the phase of radio maximum are expected to happen with a
period of $P/2\simeq 4.6$~yr.  Long term radio monitoring of the
system \citep{gregory02} shows that indeed the phase of the radio
maximum drifts from $\phi\simeq 0.45$ to $\phi\simeq 1.$ and
subsequently flips from $\phi\simeq 0.1$ to $\phi\simeq 0.45$ with
periodicity of $4.6$~yr. The radio maximum never appears within
$0.1<\phi<0.4$, in good agreement with the proposed explanation in
terms of the phase of the highest elevation of the binary orbit above
the equatorial plane of Be star disk.

\subsection{ \gr\ lightcurve.}

Electrons responsible for the TeV \gr\ emission from the system are
not able to fill the whole volume of the compactified PWN because of
the very short cooling time. The shortest cooling time is achieved at
electron energies $E_e\sim 10-100$~GeV, $t_{IC}(10^{12}\mbox{ cm},
10\mbox{ GeV})\simeq 10$~s (see Eq. (\ref{eq:tic})).  The IC
scattering on 100~GeV -- TeV electrons proceeds in the Klein-Nishina
regime. In this regime the electron cooling time grows with energy
\begin{equation}
\label{tkn}
t_{IC(KN)}\simeq 10^2\left[\frac{L_*}{10^{38}\mbox{ erg/s}}\right]
\left[\frac{\epsilon_{IC}}{1\mbox{ TeV}}\right]^{0.7}
\left[\frac{D}{10^{12}\mbox{ cm}}\right]^2\mbox{ s}
\end{equation}

Propagation of electrons with the energy about 100~GeV is affected by
 development of the pair production cascade in the field of soft
photons from Be star (with typical energy $\epsilon_*\sim
1-10$~eV). The density of such photons is
\begin{equation}
n_{ph}\sim 10^{13}\left[\frac{L_*}{10^{38}\mbox{ erg/s}}\right]
\left[\frac{10^{12}\mbox{ cm}}{D}\right]^2\mbox{ cm}^{-3}
\end{equation} 
Taking into account that the pair production cross-section peak at the value
$\sigma_{\gamma\gamma}\simeq 10^{-25}$~cm$^2$ for \gr s of energy
$E_\gamma\simeq \left[10\mbox{ eV}/\epsilon_*\right]100\mbox{ GeV}$,
one can estimate the optical depth for the 100~GeV -- 1~TeV \gr s as 
\begin{equation}
\tau_{\gamma\gamma}\simeq 1\left[
\frac{L_*}{10^{38}\mbox{ erg/s}}\right]\left[
\frac{10^{12}\mbox{ cm}}{D}\right]
\end{equation}
The life time of both 100~GeV-TeV and 10~GeV electrons is much shorter
than the escape time to the region of dominance of synchrotron loss,
$D\ge 10^{13}$~cm. The cascade electrons and positrons loose energy
mostly via IC (rather than via synchrotron) emission\footnote{For
  electrons with energies $E_e>1$~TeV the synchrotron loss time
  (\ref{eq:tsynch}) can be comparable or shorter than the IC cooling
  time (\ref{tkn}) even in the region $D<10^{12}$~cm.  TeV electrons
  emit synchrotron radiation in $100$~keV energy band (see
  Eq. (\ref{esynch})). Taking into account that luminosity of the
  system at $\sim 100$~keV energies is just by a factor of several
  higher than its TeV luminosity, one can find that a non-negligible
  synchrotron contribution can be present in the 100~keV energy band
  during the phases around the maximum of TeV lightcurve.}.  This
means that development of electromagnetic cascade efficiently channels
the power initially injected into 100~GeV -- TeV energy band to the
energy band around $\sim 10$~GeV.

Since the optical depth with respect to the pair production is highest
near the periastron, one expects to observe a "deep" in the
100~GeV -- TeV lightcurve of the source around the phases $\Phi\sim
180^\circ$ ($\phi\sim 0.3$) in agreement with MAGIC observations \citep{albert06}.
 The shape of the source lightcurve in the
energy band below $\sim 10$~GeV should be qualitatively different from
the one in the 100~GeV-TeV energy band. Since (a) the $10$~GeV
emission is not uppressed by the pair production and (b) a part of the
$10$~GeV power is generated via the development of electromagnetic
cascade in the source, the maximum of the 10~GeV flux is expected
roughly during the phases of maximal suppression of the 100~GeV-TeV
flux. In other words, future observations of the source with {\it
  AGILE} and {\it GLAST} are expected to find a maximum, rather than
minimum of the 1-10~GeV lightcurve around the phases $\Phi\sim
180^\circ (\phi\sim 0.3$).

\subsection{Model summary.}

To summarize, we find that the ``compactified PWN'' model
can explain all observational properties of \lsi, from radio to TeV
\gr\ energy band. The main assumption of this model is that the nebula
has an "onion-like" structure. The inner region of the nebula at
$D<10^{12}$~cm is characterized by the dense stellar wind and dense
photon background. In this region the 10~MeV electrons, responsible
for the radio and X-ray emission loose energy mostly on heating of the
stellar wind, while 100~GeV-TeV electrons initiate the development of
pair cascades. This explains the suppression of the radio, X-ray and
TeV \gr\ luminosity of the source close to the periastron.  The X-ray
luminosity peaks when the pulsar moves out of the dense central part
into the intermediate region $D>10^{12}$~cm where the dominant energy
loss mechanism is the IC scattering of soft photons from Be star. The
radio luminosity reaches its maximum during the phases when 10~MeV
electrons are efficiently supplied by the fast polar wind from Be star
to the outermost part of the nebula, $D>10^{13}$~cm, where the synchrotron
loss dominates.

Qualitative analysis of behaviour of the X-ray and radio lightcurves
of the system has enabled us a tentative determination of the
orientation of the equatorial plane of the wind from Be star (at the
moment of 1996 \rxte\ observations), in which the density of the
stellar wind is highest. Since the correlation of the maximum of X-ray
lightcurve with the moment of the pulsar passage through the middle of
the disk is one of the key points of our model, it is interesting to
find a confirmation for this fact in experimental data.

A complimentary information about the orientation of the binary orbit
with respect to the Be star disk can be obtained from the observation
that the passage of the pulsar from below to above (from the point of
view of observer on the Earth) the densest part of the disk is
associated with the decrease of the column density of matter between
the position of the pulsar and the observer. Passage of the pulsar
from above to below the densest part of the disk is then associated
with the rise of the column density. Taking into account that the
estimated density of the stellar wind in the most compact part of the
PWN is $n>10^{10}$~cm$^{-3}$, while the thickness of the dense layer
of the wind is about the size of the star, $\sim R_*\sim
(several)\times 10^{11}$~cm, one expects to find the variations of the
matter column density
\begin{equation}
\Delta N_H\sim nR_*\ge 10^{21}\mbox{ cm}^{-2}
\end{equation}
Such variations are, in principle, detectable by \xmm. 

As it is mentioned above, in our model the suppression of the X-ray
maximum during the pre-periastron passage of the pulsar through the
densest part of the Be star disk is explained by the dominance of the
Coulomb loss in the innermost part of the stellar wind. The dominance
of the (energy independent) Coulomb loss leads to hardening of
electron spectrum which, in turn, should lead to the hardening of IC
emission spectrum which is also a potentially detectable effect.

In the following section we make an attempt to find both the variation
of $N_H$ and the hardening of the spectrum associated to the pulsar
passage through the densest part of the disk in the available
observational data on \lsi.  The re-analysis of experimental data
allows us to further constrain the geometry of the system (in
particular, its orientation with respect to the line of sight).

\section{Re-analysis of the archive X-ray data.}

In the previously published analysis of the available X-ray data, the
hydrogen column density was fixed during the model fitting of the
data, so that no information about presence/absence of the short and
long time scale variations of $N_H$ is available. The main reason for
this is the difficulty of "disentanglement" of variations of $N_H$
from the variations of the photon index, $\Gamma$, in the commonly
accepted X-ray spectral model of the source, which consists of a
powerlaw spectrum modified at low energies by the
photo-absorption. Since the variations of $N_H$ by $(several)\times
10^{21}$~cm$^{-2}$ are only marginally detectable with such
instruments as \xmm, all previously published analyses assumed for
simplicity a constant value of $N_H$.

The set of available X-ray data considered in our re-analysis consists
of the monitoring of the source over a single orbital cycle with
\rxte\ \citep{harrison00}, five short \xmm\ 2002 observations
\citep{chernyakova06} and a long (50~ks) 2005 \xmm\ observation
\citep{sidoli06}.  Table \ref{tab:summary} gives a summary of the
observations analyzed below.

\begin{table*}
\caption{Journal of \rxte\ observations of \lsi \label{pcadata}}
\begin{tabular}{cccccc}
\hline
Date &           Exposure &Orbital&$\Gamma$&Flux(2-10 keV)&$\chi^2(dof)$ \\
     &            (ks)            &Phase&          &($\times10^{-11}$ ergs cm$^{-2}$s$^{-1}$)&   \\
\hline
1996 Mar 01 13:57-19:25&   7.9& 0.79&1.65$_{-0.12}^{+0.10}$& 1.17$_{-0.21}^{+0.24}$&16.93(23)\\     
1996 Mar 04 21:38-01:36&   8.0& 0.90&1.72$_{-0.09}^{+0.08}$& 1.49$_{-0.20}^{+0.23}$&11.61(23)\\ 
1996 Mar 07 23:33-03:28&   7.9& 0.01&1.86$_{-0.13}^{+0.12}$& 0.99$_{-0.20}^{+0.22}$&26.09(48)\\ 
1996 Mar 10 01:07-05:02&   7.7& 0.11&1.83$_{-0.12}^{+0.16}$& 0.94$_{-0.17}^{+0.28}$&14.79(23)\\ 
1996 Mar 13 04:06-08:19&   8.7& 0.23&1.55$_{-0.11}^{+0.10}$& 1.15$_{-0.19}^{+0.22}$&17.10(23)\\
1996 Mar 16 00:54-06:06&   8.3& 0.34&1.62$_{-0.08}^{+0.08}$& 1.62$_{-0.21}^{+0.22}$&25.16(48)\\ 
1996 Mar 18 10:44-15:48&   8.3& 0.43&1.56$_{-0.06}^{+0.06}$& 2.15$_{-0.20}^{+0.23}$&19.67(23)\\ 
1996 Mar 24 07:55-13:10&  10.2& 0.65&1.93$_{-0.10}^{+0.09}$& 1.17$_{-0.17}^{+0.19}$&17.20(20)\\ 
1996 Mar 26 00:56-06:46&  12.3& 0.71&1.84$_{-0.11}^{+0.10}$& 1.06$_{-0.17}^{+0.19}$&16.99(20)\\ 
1996 Mar 30 04:07-10:40&  12.9& 0.87&1.89$_{-0.11}^{+0.10}$& 0.93$_{-0.16}^{+0.17}$&16.50(20)\\ 
\hline
\end{tabular}
\label{tab:summary}
\end{table*}

\subsection{\rxte/PCA\ observations}

\begin{figure}
\includegraphics[width=9cm,angle=0]{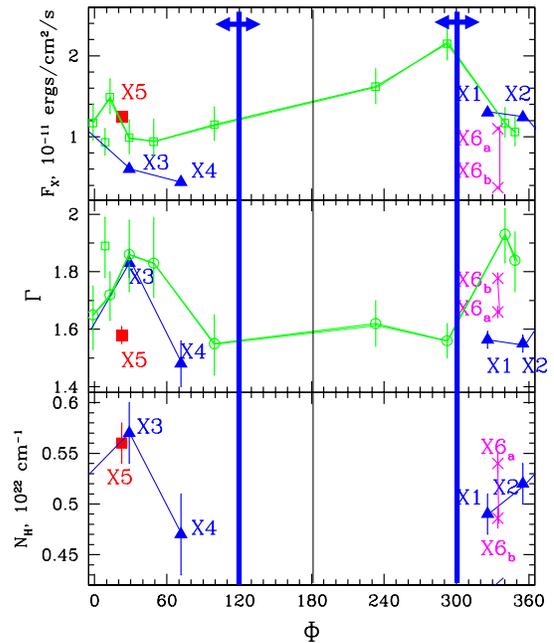}
\caption{Spectral variability of the system in 2-10 keV energy range,
  as observed with \xmm\ (red, blue and magenta points) and \rxte\
  (green open squares). Observations done during one and the same
  orbital cycle are connected with a line. The time is assumed to
  increase from left to right. Vertical thick blue lines show the suggested
  position of intersection of equatorial plane of the Be star wind
  with the binary orbit at the moment of 1996 \rxte\
  observations. Thin black vertical line marks the position of
  periastron ($\Phi$=180).}
\label{lc}										       
\end{figure}

In 1996 \rxte\ has closely monitored \lsi\, making 11 observations
along the single orbit, see Table \ref{pcadata}. For the first time
these data were presented in the paper of \cite{harrison00} where the
authors report that no significant variability of the spectral shape
is visible in \rxte\ data. Our reanalysis of the \rxte\ data with the
latest LHEASOFT 5.2 provided by the \rxte\ Guest Observer Facility
reveals, to the contrary, a systematic variation of the spectrum along
the orbit. Namely, the rise of X-ray flux is associated to the
hardening of the spectrum with the photon index decreasing from
$\Gamma\simeq 1.9$ down to $\Gamma\simeq 1.5$.

The results of our spectral analysis performed with XSPEC v 11.3.2 are
given in Table \ref{pcadata}, and are graphically represented in the
Fig.~\ref{lc}.  The  line of intersection of the equatorial plane
of the Be star disk with the orbital plane (at the moment of \rxte\
observations), suggested from the qualitative considerations of the
previous section, is shown by the vertical blue lines shifted by
$\Delta\Phi=180^\circ$. The orientation of the disk is chosen in such
a way that the maximum of \rxte\ lightcurve at $\Phi_{X1}\simeq
300^\circ$ corresponds to the moment of the pulsar passage through the
equatorial plane of the disk (in general, the disk is
geometrically thick, so the pulsar can spend most of its orbit inside
the disk). As it is discussed in the previous section, the second
maximum, at $\Phi_{X2}\simeq 120^\circ$ is suppressed because  the
pulsar passes closer to the Be star where the disk is so dense that
most of the power of high-energy electrons injected in the disk goes
onto Coulomb heating of the disk, rather than on the IC X-ray
emission. From the middle panel of Fig. \ref{lc} one can
clearly see that the suggested moment of the pre-periastron
passage of the equatorial plane at $\Phi_{X2}\simeq 120^\circ$ is
associated with the hardening of the X-ray spectrum of the system.

Unfortunately, the PCA data do not allow one to find unambiguously the
true value of hydrogen column density $N_H$, because of the lack of
sensitivity at the energies below 3~keV.  Only the information about
the variations of photon index, $\Gamma$, is available. For the
purpose of the model fit we have fixed the value $N_H=0.49\times
10^{22}$~cm$^{-2}$ derived by \citet{chernyakova06} from 2002 \xmm\
data.

\subsection{\xmm\ observations}

\xmm\ has observed \lsi\ with the EPIC instruments five times during
2002, and once in 2005. Four 2002 observations have been done during
the same orbital cycle, and the fifth one has been done seven months
later. The log of the \xmm\ data discussed in this paper is presented
in Table~\ref{data}.  These data have already been analyzed in papers
of \citet{chernyakova06} and \citet{sidoli06}. In these works it was
shown that simple power law with photoelectric absorption describes
the spectrum of \lsi\ well, with no evidence for any line features.

\begin{table}
\caption{Journal of \xmm\ observations of \lsi \label{data}}
\begin{tabular}{c@{\,}c@{\,}ccc@{\,}c}
  \hline
  Data& Observational& Date &    MJD&  Orbital&Exposure (ks)\\
  Set&   ID         &      & (days)&    Phase  &      \\
  \hline
  X1& 0112430101& 2002-02-05&52310  & 0.55  &6.4  \\
  X2& 0112430102& 2002-02-10&52315  & 0.76  &6.4  \\
  X3& 0112430103& 2002-02-17&52322  & 0.01  &6.4  \\
  X4& 0112430201& 2002-02-21&52326  & 0.18  &7.5  \\
  X5& 0112430401& 2002-09-16&52533  & 0.97  &6.4  \\
  X6& 0207260101& 2005-01-27&53397  & 0.61  &48.7\\ 
  \hline
\end{tabular}
\end{table}

The \xmmsp Observation Data Files (ODFs) were obtained from the
on-line Science
Archive\footnote{http://xmm.vilspa.esa.es/external/xmm\_data\_acc/xsa/index.shtml};
the data were then processed and the event-lists filtered using {\sc
  xmmselect} within the Science Analysis Software ({\sc sas})
v6.0.1. We re-analyzed the \xmm\ observations using the methods
described in \citet{chernyakova06}.

\begin{table}
\caption{Spectral parameters for  \xmm\ Observations of \lsi.$^*$}
\label{summary_xmm}
\begin{center}
 \begin{tabular}{l|c|c|c|c}
   \hline
   Data  & $F$(2-10 keV) &$\Gamma$&$N_H$                &$\chi^2$ (dof) \\
   Set   &$10^{-11}$erg s$^{-1}$& &($10^{22}$ cm$^{-2}$)& \\
   \hline
   X1& 1.30$\pm 0.02$ &1.60$\pm 0.03$&0.49$\pm 0.02$& 261(264) \\ 
   X2& 1.24$\pm 0.03$ &1.55$\pm 0.03$&0.52$\pm 0.02$& 272(252) \\
   X3& 0.60$\pm 0.03$ &1.83$\pm 0.05$&0.57$\pm 0.03$& 182(165) \\
   X4& 0.44$\pm 0.03$ &1.48$\pm 0.08$&0.47$\pm 0.04$&  70(76) \\
   X5& 1.25$\pm 0.03$ &1.58$\pm 0.03$&0.56$\pm 0.02$& 277(284) \\
   X6a&1.10$\pm 0.02$ &1.66$\pm 0.02$&0.54$\pm 0.01$& 653(649) \\
   X6b&0.37$\pm 0.01$ &1.78$\pm 0.02$&0.49$\pm 0.01$& 435(458) \\
   \hline
\end{tabular}
\end{center}
$^*$ Given errors represent 1$\sigma$(68.3\%) confidence interval 
uncertainties.
\end{table}

\begin{figure}
\includegraphics[width=9cm,angle=0]{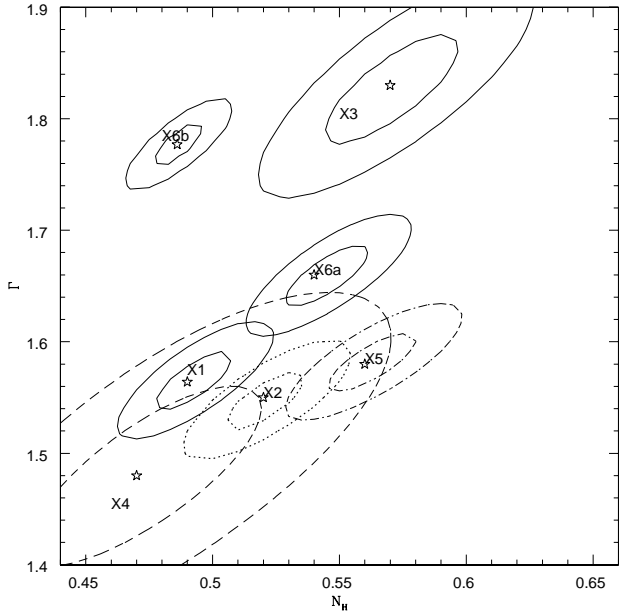}
\caption{Confidence contour plots of the column density $N_H$ and of
  the photon spectral index $\Gamma$ uncertainties for a power-law fit
  to \xmmsp observations. The contours give 1$\sigma$ (68.3\%) and
  2$\sigma$ (95.4\%) confidence levels.
\label{cntr}}										       
\end{figure}

The results of our spectral analysis are given in Table
\ref{summary_xmm} and Fig.~\ref{lc}.  Following the work of
\citet{sidoli06} we have split observation X6 into two, X6$_a$ and
X6$_b$. X6$_a$ corresponds to the first 20 ks of the observations, and
X6$_b$ to the rest.

\subsection{Systematic variation of X-ray spectrum along the orbit.}

From the bottom panel of Fig.~\ref{lc} one can see a hint of
systematic variation of $N_H$ over the orbit (the probability of
non-variable $N_H$ is 0.02\%).  From observations X1 -- X5 
one can guess that the column density appears to be lower
around the orbital phases closer to periastron and higher at apastron. 
At the same time, the variation of $N_H$ at 50~ks scale 
during the observation X6 is comparable to the scatter of $N_H$ values on the orbital time scale.
Obviously, more observations, especially  around the periastron, $120^\circ<\Phi<240^\circ$ 
would help to clarify the question of the systematic orbital modulation of equivalent hydrogen column density.

A possible interpretation of the orbital modulation of $N_H$ 
is that the part of the binary orbit around
the periastron is situated closer to observer. In this case the part of the orbit around the apastron is embedded 
"deeper" into the stellar wind, from the viewpoint of observer and photons have to pass through a higher matter column density.

The middle panel of Fig. \ref{lc} shows the orbital evolution of the
photon index $\Gamma$. One can see that a spectral hardening similar
to the one found in \rxte\ data is observed at the phases $\Phi\sim
120^\circ$ (suggested pre-periastron passage of the equatorial plane
of the disk).  However, contrary to the \rxte\ observations, no
softening of the spectrum is observed after the phase $\Phi_{X1}\simeq
300^\circ$. In general, a large scatter of the values of X-ray flux
and photon indexes are found at the phases $\Phi>300^\circ$.  This can
be explained either by physical arguments (close to the apastron the
density of clumps in the stellar wind decreases which leads to the
significant variability of the flux and the spectral index), or,
possibly, by the low quality of X-ray data (the precision of the
measurements can be insufficient to measure independently $N_H$ and
$\Gamma$).

To address this question in more details, and to avoid the possible
degeneracy on the determination of the photon index and column density
we show in Fig.~\ref{cntr} the 1$\sigma$ and 2$\sigma$ confidence
levels contour plots of spectral index $\Gamma$ versus $N_H$. From
this figure one can see that un-freezing of $N_H$ in the model fit
does not prevent the detection of variability of $\Gamma$ over the
orbit.

\subsection{Short time scale variations of $N_H$.}

The long \xmm\ observation of the system in 2005 as well as radio
observations by \citet{peracaula97} reveal the variability of the flux
from the system on the time scales of several
kiloseconds. Interpretation of the short time scale variability in the
model of interaction of the pulsar wind with the clumps of the stellar
wind suggests that both the photon index $\Gamma$ and the column
density $N_H$ should vary at this time scale.

\begin{figure}
\includegraphics[width=9cm,angle=0]{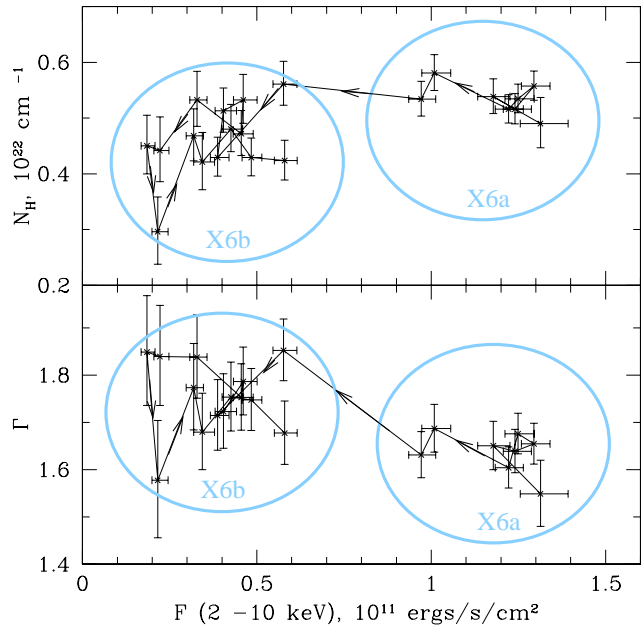}
\caption{ Column density (top panel) and spectral slope (bottom panel) 
dependence on the 2 - 10 keV Flux during the X6 observation
 with a bin time of 2000 s. Arrows show the direction of time evolution.
 \label{set6}}
\end{figure}

To study the possible short time scale variability of both $N_H$ and
$\Gamma$ we have split the long \xmm\ observation onto time bins of
duration of 2~ks, extracted the spectrum of the source separately in
each time bin and fitted it with the absorbed powerlaw model. The
resulting dependence of the column density $N_H$ and the photon index
$\Gamma$ on the flux from the source is shown in top and bottom panels
of Fig.~\ref{set6}.

In the top panel of Fig. \ref{set6} one can see that the column
density decreases when the flux drops. The probability that $N_H$ was
constant during the whole 50~ks observation is less than 0.03\%. Such
behaviour is natural to expect within the "clumpy wind" model: lower
density of the clumps and/or smaller number of active
(i.e. illuminated by the pulsar wind) clumps should result in the
lower luminosity of the system.

Bottom panel of Fig.~\ref{set6} shows the dependence of the spectral
index $\Gamma$ on the 2 - 10 keV flux. Softening of the spectrum is
observed during the period of a lower flux values. Such behaviour is
also expected in the "clumpy wind" model: cooling of high-energy
electrons at the end of activity of a clump leads to the softening of
the spectrum of IC emission, as it is discussed in
\cite{chernyakova06}.

\subsection{High and low flux spectral states of the source.}

 The direction of the time evolution  in  Fig.~\ref{set6}
is shown with arrows. In both panels data points clearly split into
two groups, corresponding to X6a (group of points on the right with
high flux), and X6b (group of points on the left with low flux)
observations on the Fig.~\ref{lc}. The qualitative behaviour of the
system is somewhat different in the "high flux" and "low flux"
state. In the "high flux" state the characteristics of the system
"stabilize", in the sense that a well defined value of the photon
index, close to $\Gamma\simeq 1.5$, and of $N_H$, close to $N_H\simeq
0.5\times 10^{22}$~cm$^{-2}$, are found. To the contrary, in the "low
flux" state the systems exhibits an irregular behaviour, with
significantly variable $N_H$ and $\Gamma$.

\begin{figure}
\includegraphics[width=9cm,angle=0]{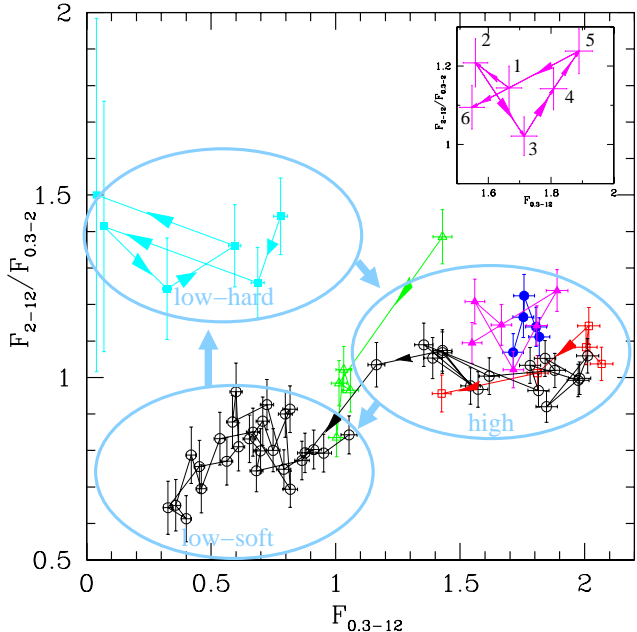}

\caption{ Hardness ratio versus count rate in the 0.3-12 keV energy
  range for all \xmm\ observations (EPIC pn) of the source, with a bin
  time of 1000 s. The hardness ratio is defined as ratio between
  counts in the hard 2-12 keV and soft 0.3-2 keV energy range. Open
  red squares correspond to observation X1, filled blue circles to X2,
  open green triangles to X3, cyan filed squares to X4, filled magenta
  triangles to X5, and open black circles to X6. Arrows show the
  direction of time evolution. The enlarged view of the X5 observation
  is shown on the inset.}
\label{hardn}
\end{figure}

To test the distinction between the low and high flux states we have
investigated the spectral evolution of the system on the 1~ks time
scale also in the 5 short \xmm\ observations. Fig. \ref{hardn} shows
the results of our analysis. To make the definition of the "spectral
state" of the system independent of the model used to fit the
spectrum, we plot in this figure the model-independent "hardness
ratio", as a function of the X-ray flux. One can see that indeed, in
the "high flux" state, the hardness ratio does not exhibit significant
variations, which signifies that the spectrum of the system is
stable. To the contrary, the "low flux" states can be divided onto
"low soft" and "low hard" states, corresponding to the different
values of the hardness ratio.

The pattern of the source variability in the "hardness ratio vs. flux"
diagram turns out to be qualitatively different from the pattern
observed in conventional accreting X-ray binaries \citep{zdziarski04}.
Namely, the spectral evolution proceeds along the "high-hard"
$\rightarrow$ "intermediate-soft" $\rightarrow$ "low-soft"
$\rightarrow$ "low-hard" $\rightarrow$ "high-hard" loop. This loop is
exactly opposite to the "low-hard" $\rightarrow$ "intermediate-soft"
$\rightarrow$ "high-soft" $\rightarrow$ "intermediate-hard"
$\rightarrow$ "low-hard" evolution loop observed in accretion powered
X-ray binaries.

The observed evolution pattern in the hardness ratio vs. flux diagram
can be naturally explained in the compactified PWN model of the
source. A stationary spectrum of the "high" state is achieved when a
sufficiently large number of the stellar wind clumps is illuminated by
the pulsar wind. Drop of injection of the high-energy electrons at the
end of activity of the clumps leads to the softening of the X-ray
spectrum during the flux decrease. Hard spectrum in the low flux state
is achieved during the pulsar passage through the densest part of the
stellar wind.

\subsection{Mini-flares produced by activity of individual clumps.}

The short, kilosecond time scale, variability of $N_H$ can be
associated with the variations of the density of the stellar wind from
clump to clump. Indeed, if the typical density of the clump is $n\sim
10^{10}$~cm$^{-3}$, while its size is $R_{clump}\sim 10^{11}$~cm, one
expects to find the fluctuations of the column density $\delta N_H\sim
nR_{clump}\sim 10^{21}$~cm$^{-2}$ when the activity of individual
clumps switches on/off.

\begin{figure}
\includegraphics[width=9cm,angle=0]{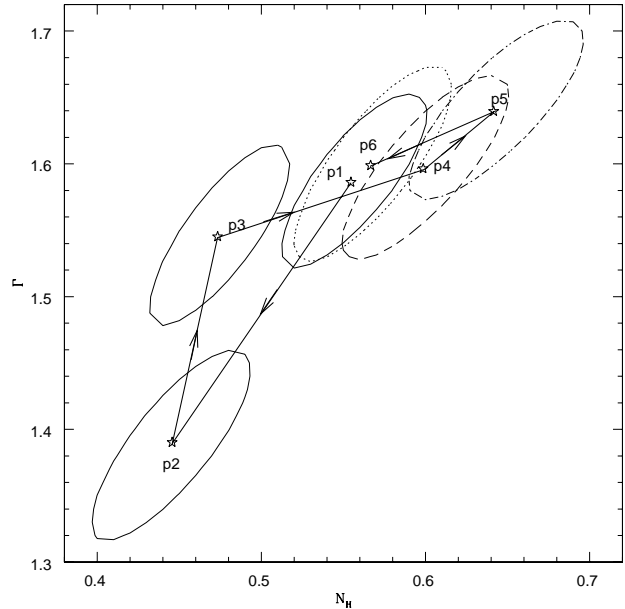}
\caption{ Confidence contour plots ( 1$\sigma$ level) of the column
  density $N_H$ and of the photon spectral index $\Gamma$
  uncertainties for a power-law fit to each 1000 s of X5
  observation. }
\label{cntr5}
\end{figure}

The activity of individual clumps results in the short, several
kilosecond, "mini-flares" observed in the X-ray data. Examples of such
"mini-flares" are visible e.g. in the \xmm\ observations X6b and X5
which is shown in more details in the inset in the right top corner of
Fig.  \ref{hardn} and in Fig. \ref{cntr5}. Similarly to the case of
the long \xmm\ observation, we split the whole observation X5 into 6
intervals of duration of 1~ks and extract the spectra of the source
separately in each interval.  One can see that the X-ray flux started
to grow during the second ks interval, reached its maximum in the 5th
interval and subsequently dropped to its initial value during the
interval 6. The column density $N_H$ started to grow exactly at the
moment of the on-set of the mini-flare, during the interval 2 and
continued to grow until the flux reached its highest value during
interval 5. The scale of the variation of $N_H$ during the mini-flare
enables to estimate the column density of the clump responsible for
the flare, $N_{H,clump}\simeq 10^{21}$~cm$^{-2}$. Remarkably, the
spectrum of the source hardened on the time scale of 1~ks at the
moment of the onset of the mini-flare (period 2) and stabilized at the
value $\Gamma\simeq 1.6$ afterwards. Such transient hardening of the
spectrum on the time scale shorter than the inverse Compton cooling
time (\ref{eq:tic}) can be explained by the fact that the equilibrium
electron spectrum determined by the balance of injection and cooling
rates is established only at the time scales of about $t_{IC}$
(\ref{eq:tic}).

\section{Conclusions.}

We have developed a model of the "compactified" pulsar wind nebula of
the \gr -loud binary system \lsi. It turns out that this model can
explain the observed properties of the source in radio, X-ray and \gr\
energy bands.  In this model the maximum of the X-ray flux from the
source is reached when pulsar crosses the middle plane of the dense
equatorial wind from Be star.  Maximum of the radio emission is
achived when the pulsar wind penetrates most efficiently into the
region of fast polar wind from Be star. Maximal luminosity in
100~GeV-TeV energy band is achieved when the propagation of very-high
energy \gr s is least affected by the pair production.  The naively
expected maxima of X-ray and  radio lightcurves close to
the phase of periastron of the binary orbit are suppressed because of
the dominance of Coulomb loss inside the densest part of the stellar
wind.

Association of the observed X-ray maximum with the period of the
pulsar passage through the densest part of the disk has enabled us to
constrain the orientation of the binary orbit with respect to the
anisotropic wind from Be star.  To test the conjecture about the
orientation of the binary orbit we have searched for the variations of
the hydrogen column density (which can be associated with the pulsar
passage through the dense part of the stellar wind) in the available
X-ray data. We find a marginally detected orbital variation of $N_H$.  
This orbital variation of
$N_H$, if confirmed with more X-ray observations around the
periastron, constrains also the orientation of the system with respect
to the line of sight.

Our re-analysis of available X-ray observations of \lsi\ has also
revealed that the changes in the spectral state of the source are
accompanied by the changes in the hydrogen column density by $\Delta
N_H\sim 10^{21}$~cm$^{-2}$, not only on the long (orbital) time scale,
but also on the short (several kilosecond) time scale. This supports
the hypothesis that the short, kilosecond time scale variations of the
source flux are caused by the motion of compact object through the
dense clumps of wind from the companion Be star.
 
\section{Acknowledgments}

The authors would like to thank M. Revnivtsev for the help with \rxte\ data
analysis. We are also grateful to V.Bosch-Ramon for useful discussions of the
subject.

 \label{lastpage}

\end{document}